\documentclass{elsart}

\usepackage{graphics}
\usepackage{graphicx}

\usepackage{amssymb}
\usepackage{amsmath}

\begin{document}

\begin{frontmatter}



\title{Detection of Noble Gas Scintillation Light with Large Area Avalanche Photodiodes (LAAPDs)}


\author{R. Chandrasekharan}
\author{M. Messina}
\author{A. Rubbia}
\address{Institut f\"{u}r Teilchenphysik, ETHZ, \\ CH-8093 Z\"{u}rich,
Switzerland}
\begin{abstract}
Large Area Avalanche Photodiodes (LAAPDs) were used for a series of
systematic measurements of the scintillation light in Ar, Kr, and Xe gas.
Absolute quantum efficiencies are derived. Values for Xe and Kr are
consistent with those given by the manufacturer. For the first time we
show that argon scintillation (128 nm) can be detected at a quantum
efficiency above 40$\%$. Low-pressure argon gas is shown to emit
significant amounts of non-UV radiation. The average energy
expenditure for the creation of non-UV photons in argon gas at this
pressure is measured to be below 378 eV.

\end{abstract}

\end{frontmatter}

\section{Introduction}
Noble gases are known to provide scintillation light with high yield, 
comparable to that of NaI(Tl). Detection of this light proves fundamental
in many applications in which noble gas is used as medium (for example
in TPCs the prompt scintillation light is used for triggering and $T_0$ determination). Difficulties arise from the fact that noble gas emission is peaked in the deep ultraviolet (DUV) range. To detect such light, wavelength shifter-coated Photomultiplier Tubes (PMTs) have often been used, resulting in a very low global quantum efficiency. There is interest in knowing if large area
avalanche photo-diodes (LAAPDs) could be a viable alternative, 
in particular in applications where the radiopurity of the PMT's glass
is a concern (e.g. direct dark matter searches, detection of solar
neutrinos, etc.). 

The results presented in this paper show that it is in principle possible
to detect gas scintillation from Kr, Xe and Ar with APDs with
a quantum efficiency higher than with PMTs.
The issues related to the signal to noise of APDs, to the parallel 
operation of a large number of APDs to increase
the sensitive area and the mechanical problems at potential cryogenic
temperatures remain to be solved.

  \section{Noble Gas Scintillation}
  \label{sec:nobgas}
Gas excitation and ionization can lead to the emission of
scintillation photons in the DUV range
via the processes \cite{doke89} of
excitation
\begin{equation}
\begin{split}
R^*+ R & \rightarrow R^*_2\\
 R^*_2 & \rightarrow 2R + h \nu,\\
\end{split}
\end{equation}
and ionization
\begin{equation}
\begin{split}
e_{hot}+ (collisions)  &\rightarrow e_{th}\\
R^+ + R  &\rightarrow R^+_2\\
 R^+_2 +e_{th}  &\rightarrow R^{**} + R\\
 R^{**} & \rightarrow R^* + \rm heat\\
R^* + R  & \rightarrow R^*_2\\
 R^*_2 & \rightarrow 2R + h\nu.\\
\end{split}
\end{equation}
The possible emission of non-UV light, mostly in the IR region, 
is attributed to the transitions between excited atomic states. 
Indeed, the atomic spectrum of argon shows several very intense lines at
wavelength 700-1050 nm. A set of lines located between 400-500nm has
intensities below 3 $\%$ of the most intense lines found in the
infrared region.

According to \cite{hurst,miyajima,platzman,icru}, the
average energy expended per ion pair $W_g^{ion}$ can be related to the
ionization potential $I$
\begin{equation}
\label{eq:Wgi}
\frac{W_g^{ion}}{I}=\frac{E_i}{I} + \left(\frac{E_{ex}}{I}\right)\left(\frac{N_{ex}}{N_i}\right)+ \frac{\epsilon}{I},
\end{equation}
where $N_i$ is the number of ions produced at an average energy
expenditure of $E_i$, $N_{ex}$ is the number of excited atoms produced
at an average expenditure of $E_{ex}$, and $\epsilon$ is the average
kinetic energy of sub-excitation electrons.  Equation \ref{eq:Wgi} can
also be applied to condensed noble gas if $I$ is substituted by the
band-gap energy $E_g$. Values used for calculations in this work
are shown in Table~\ref{tab:ionization}.

The energy balance equation is energy dependent in all four every
terms, however, for $E \gg I$ this dependence is weak. For
$\alpha$-particles in argon, $W_g^{ion} = 26.5\pm 0.5$ eV at energies
$E_\alpha \ge 1$ MeV. For $E_\alpha = 0.1$ MeV, the value is only
somewhat higher at $W_g^{ion} = 27.5\pm 1.0$, increasing further as the
kinetic energy is reduced. Our measurements were performed in a
pressure range where the scintillation is brought forth by $\alpha$
particles with at least 0.5 MeV kinetic energy.

\begin{table}[htb]
\centering
\begin{tabular}{|c|c|c|c|c|c|c|c|}
\hline
  & $W_g^{ion}$ & $N_{ex}/N_i$ & I & $\epsilon/I$   & $E_{i}/I$ & $E_{ex}/I$ & UV Peak Wavelength\\
\hline
Ar & 26.34 & 0.4 & 15.7 & 0.33 & 1.06 & 0.85 & 128 nm\\
Kr & 24.1  & 0.4 & 13.9 & 0.36 & 1.06 & 0.85 & 150 nm\\
Xe & 21.9  & 0.4 & 12.1 & 0.39 & 1.1  & 0.85 & 175 nm \\
\hline
\end{tabular}
\vspace{0.5cm}
\caption{Values used for calculations in this work. Energies are
expressed in eV. From \cite{hurst,miyajima,platzman,icru}. }
\label{tab:ionization}
\end{table}

Assuming no ionization contribution to UV scintillation light, justified at
the low pressures used in this work~\cite{suzuki,carvalho}, the average energy expenditure
per photon is
\begin{equation}
\label{eq:Wg}
W_\gamma^{DUV} = \left(\frac{N_i}{N_{ex}}\right)\cdot E_i + E_{ex} +\left(\frac{N_i}{N_{ex}}\right)\cdot \epsilon
\end{equation}
electron volts. This yields $W_\gamma^{DUV}=67.9$ eV, $61.2$ eV, and $55.9$ eV
for Ar, Kr, and Xe, respectively.

  \section{Experimental Set-Up}

\begin{figure} 
\includegraphics[width=\textwidth]{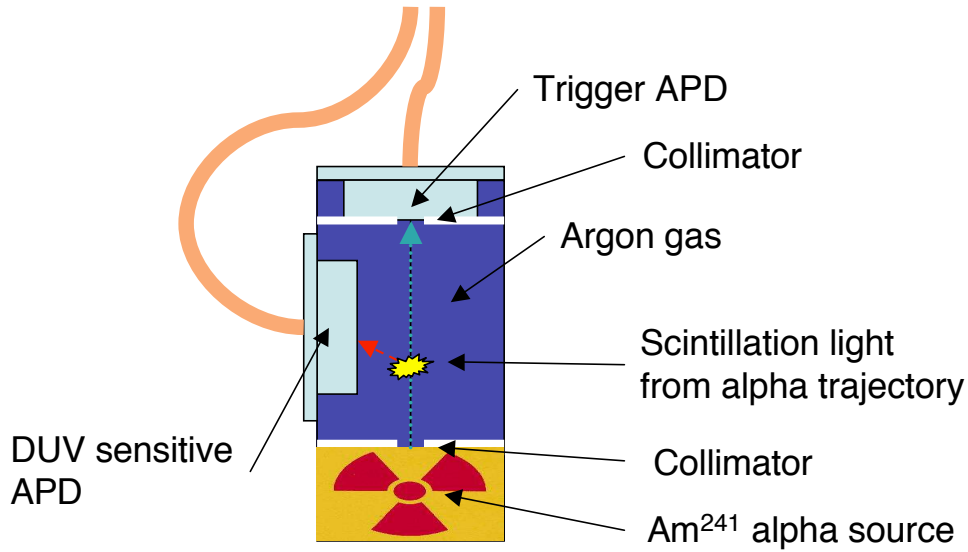} 
\caption{Alpha particles pass from the source to an APD serving as
trigger. A laterally mounted APD measures the scintillation light
emitted along the trajectory.}
\label{fig:setup}
\end{figure}

In our experimental setup (See Figure~\ref{fig:setup}), we used
an $^{241}$Am source which emits $\alpha$-particles of an initial
energy of 5.486 MeV (85\%) and 5.443 MeV (13\%).  For the
measurements, $\alpha$-particles pass from an open $^{241}$Am source
to an APD employed as a trigger. The triggered trajectories
necessarily pass through the noble gas, causing scintillation. An APD
mounted on an axis perpendicular to the trajectory detects this
light. The set of trajectories are contained within a cylindrical
region of 1.5mm radius and 45mm length. The trigger is simply an APD
operated in unitary gain mode.

The APDs are Advanced Photonix LAAPDs with an active diameter of
16mm \cite{advphot}. For scintillation light detection a windowless, DUV-enhanced device was
used primarily, cross-checked against a windowless Red/IR-enhanced
device.

\begin{figure}[htb] 
\begin{centering}
\includegraphics[width=.6\textwidth]{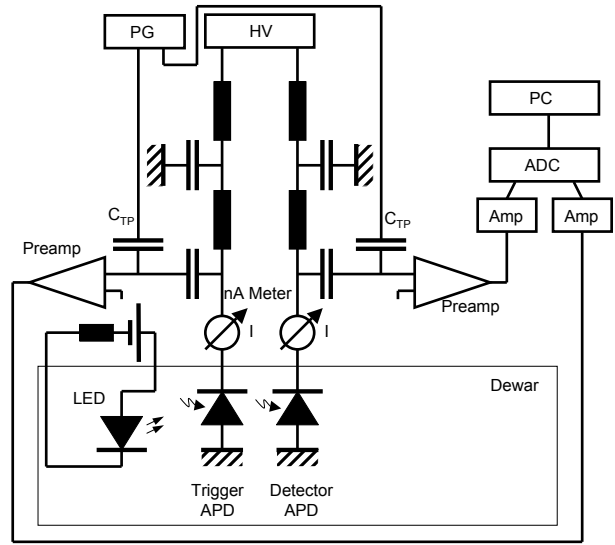} 
\caption{\label{fig:schem} Schematic of the electronics including LED
and ampere meters used for calibration. Labeled components are
described in text. Trigger and detector APDs have almost identical
electronics chains. }
\end{centering}
\end{figure}

To perform measurements, the dewar is evacuated to about $10^{-5}$
mbar, flushed a few times, and filled with the noble gas in which the
measurement is to be performed. For the measurements in argon, the gas
is purified by an Oxysorb cartridge.  The condition that the $\alpha$
deposits at least 0.5 MeV in the trigger imposes a gas-dependent upper
limit on the pressures at which the measurements could be
performed. A lower bound is given by the electrical discharge-tendency
encountered at low pressures. External cooling allows a gas
temperature range from $-5$ to 20~degrees~C. Thermistors monitor the
temperature of the gas as well as of the APD, the latter being of
importance due to the temperature dependence of APD gain.

A schematic of the electronics is shown in Figure~\ref{fig:schem}.
Directly outside the dewar, the APD signals are decoupled and
amplified using a ICARUS charge-sensitive hybrid preamplifier \cite{centro}. The
feedback capacitance of the preamplifier is modified to $5.7$ pF in order to be
better adapted to the large capacitance of the APD. The preamplified signal is
shaped and amplified by a Canberra 2020 Spectroscopy Amplifier.

The signal height is obtained by comparison with a test pulse,
consisting of a pulse generator injecting a charge via
a capacitor of known capacitance. Measuring the current flowing through the APD allows the
gain to be monitored according to the method described in
\cite{Karar:1999pg}, where a LED situated inside the dewar serves as a
continuous light source.

\begin{figure}[htb]
\includegraphics[height=.25\textheight,width=0.5\textwidth]{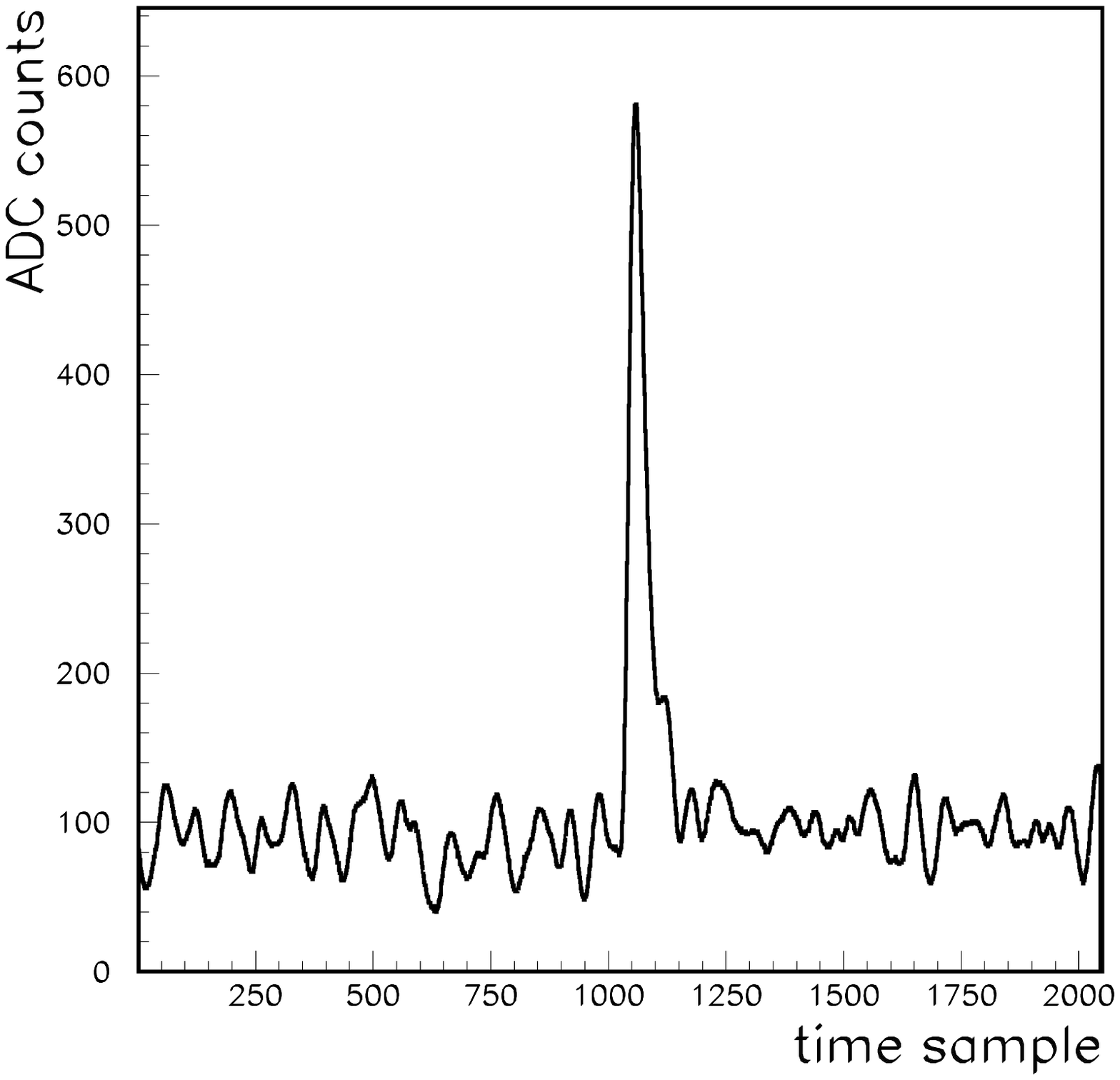}
\includegraphics[height=.25\textheight,width=0.5\textwidth]{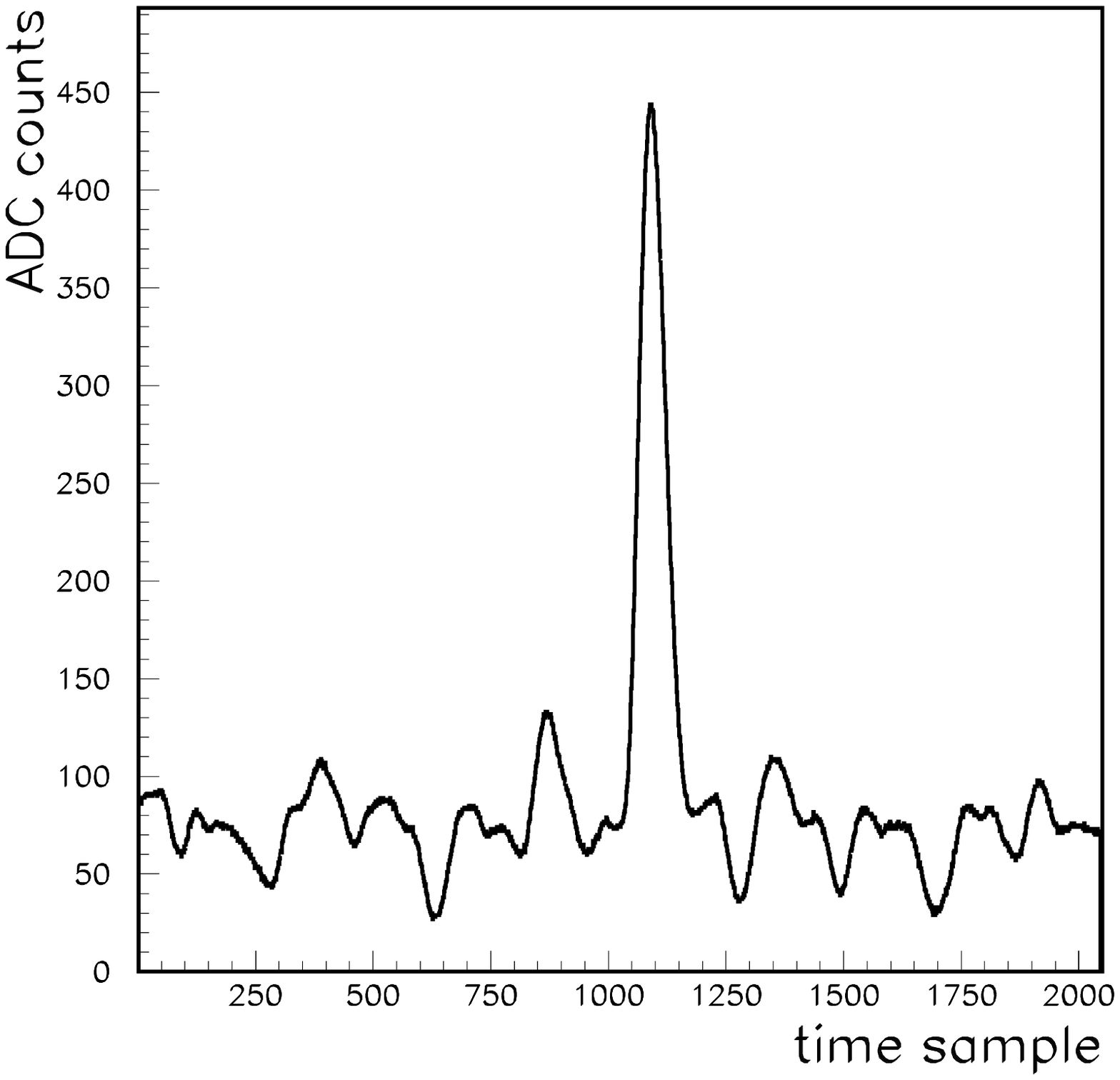}
\caption{ Plot shows typical signals with 5 $\mu$s (left) and 10
$\mu$s shaping time (right). Time samples correspond to 400 ns
each. The signals were randomly picked from the data taken in argon at
0.783 atm and 0.777 atm respectively, see Table \ref{tab:ar} for more
information about the conditions under which the data was acquired. }
\label{ec2_sum}
\end{figure}

Typical signals are shown in Figure~\ref{ec2_sum}.

\section{Calibration}

The raw charge $Q_S$ of measured signals is obtained by comparison with a 
test pulse
\begin{equation}
\label{eq:c1}
Q_{S} = \frac{S_S}{S_{TP}}\cdot Q_{TP} 
\end{equation}
where $S_{S}$, $S_{TP}$ are the peak heights of the signal and the test pulses,
respectively. The precision of the measurement depends crucially on the
knowledge of the amount of charge
$Q_{TP} = V_{TP}\cdot C_{TP} $
that the test pulse injects into the preamplifier. A pulse generator produces step
functions of height $V_{TP}$. $V_{TP}$ is on the scale of several millivolts
and can be determined with a precision better than $2\%$. $C_{TP}$ is of the
order of 5 pF. Its value can be measured more precisely by comparing
the test pulse with the charge signal $Q_{S_\alpha}$ of the 5.486
MeV $\alpha$ ionization in the trigger APD under vacuum conditions. Since
the energy $W_{Si}$ necessary to produce an electron-hole pair in the silicon
of the APD is known to be 3.65 $\pm$0.05 eV \cite{aprile,icru},
\begin{equation}
\frac{Q_{S_\alpha}}{e} = \frac{E_\alpha}{W_{Si}}|_{vac}  
\end{equation}
where $e$ is the elementary charge. In this way,
a value $C_{TP} = 5.04 \pm 0.16 \, \rm pF$
was measured.

 \section{Estimation of Absolute Photon Yield}
\label{CoPY}

$W_\gamma^{DUV}$, the amount of energy necessary for the creation of a
DUV scintillation photon in the given gas can be calculated from
Equation \ref{eq:Wg} (See Section~\ref{sec:nobgas}).  Dividing the
$\alpha$-particle's total energy loss in gas by this number gives an
\it upper limit \rm to the number of DUV photons. Quenching, or any
unaccounted degree of freedom, leads to a reduction of the number of
photons.

In order to correlate results from different gases, we have
developed a simulation to estimate the absolute photon
yield in our setup. When the $\alpha$-particles pass through the
noble gas of known temperature and pressure, their energy loss along
the well defined trajectory is given by \cite{NIST}. We have cross-checked
our calculations
of the energy loss of the $\alpha$ particle by
increasing the gas pressure to the threshold pressure above which the
$\alpha$ loses all its energy before reaching the trigger. The
calculation accurately predicts this value. The simulation assumes
trajectories on straight lines. This is justified since the detour
factor, defined as the ratio of the projected range to the actual
length of the $\alpha$ trajectory calculated in the
continuous-slowing-down-approximation, is 0.98 for $\alpha$ particles
of energies around 5.5 MeV \cite{NIST}. By requiring a minimal energy
deposition of 0.5 MeV in the trigger, this uncertainty is further
reduced.

The detector's solid angle can be calculated for each point along the
trajectory. For calculation, two models were used to simulate photon
emission. The first assumes isotropic emission of photons at each
point of the $\alpha$'s trajectory. The second model assumes emission
of photons in a plane perpendicular to the trajectory as suggested by
\cite{moerman}. Both gave comparable numbers, $N_{iso}$ being
approximately $15\%$ smaller than $N_{perp}$. Since the model of
isotropic emission fits our xenon and krypton data better, $N_{iso}$
was assumed to be the better model. In the following, $N_{iso}$ will
just be referred to as $N_{\gamma}$.

An estimate of the non-UV light yield is more difficult. Although
infrared emission has been detected in argon and xenon
\cite{bressi,belogurov}, quantitative data is hard to find
in the literature. In the case of argon, we have estimated
the non-UV contribution (See Section~\ref{IRC}) and
checked the effect on our estimation of the quantum efficiency.

  \section{Quantum Efficiency}

\begin{figure}[h] 
\includegraphics[width=\textwidth]{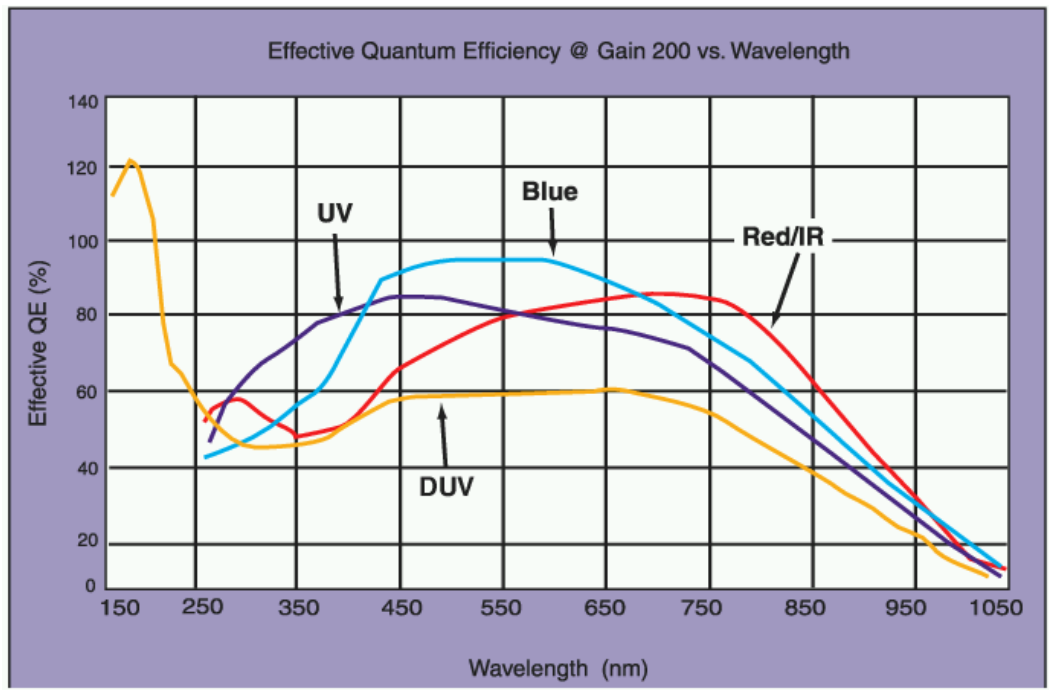} 
\caption{\label{APIqe} The quantum efficiency of the LAAPDs \cite{advphot}. The DUV
enhanced device used in our measurements has a high
quantum efficiency in the region of xenon scintillation light.}
\end{figure}

The external quantum efficiency of APDs, from now on referred to as
quantum efficiency, is defined as the number of primary electron-hole
pairs produced per incident photon. The quantum efficiency
$\epsilon_Q(\lambda)$ is a function of the wavelength of the incident
light. In Figure \ref{APIqe}, the quantum efficiency of the used LAAPD
is given by the manufacturer \cite{advphot}.

To our knowledge, measurements of the quantum efficiency of these
devices for wavelengths below 150nm, specifically at the 128nm of
argon scintillation light, have not been published.  As a
reference, the highest values for the global quantum efficiency of
\emph{PMTs} in argon scintillation light detection were obtained by
\cite{Benetti}.  The measurements were made using a TPB wavelength
shifting-coating, resulting in quantum efficiency values below 10
$\%$.  It is therefore quite relevant to understand the quantum
efficiency of the APD also in this region of the spectrum.

\section{Measurements}

The quantum efficiency can be obtained with
\begin{equation}
\label{eq:eqcal}
\epsilon_{q0}(128nm) = \frac{Q_S }{(e\cdot N_\gamma\cdot G)} ,
\end{equation}
where $e$ is the elementary charge, $N_\gamma $ is the number of
photons impinging on the LAAPD as predicted by our simulation
(See Section \ref{CoPY}), and $G$ is the measured APD
gain. 

Equation \ref{eq:eqcal} is exact for pure argon gas with emission only
at 128 nm detected by an idealized detector. Possible absorption effects due to
impurities in the gas lead to efficiency under-estimation. Further,
the employed shaping time needs to be sufficient to integrate over the
full characteristic light emission time. Using an inferior shaping time leads to efficiency
under-estimation. Competing with this effect is the non-UV
contamination. The DUV-enhanced APD has a non-zero quantum efficiency
throughout the visible and in parts of the IR region, see Figure
\ref{APIqe}. Therefore, any unaccounted non-UV contribution leads to
efficiency over-estimation. In this work, the measurement of the
non-UV photon contribution was performed only in argon.

To verify the correctness of our photon yield simulation, measurements
were taken in krypton and xenon using DUV-enhanced LAAPD. These
devices' quantum efficiencies are given by the manufacturer at the
wavelengths of Kr and Xe DUV-emission. Tables \ref{tab:krxe} and
\ref{tab:xe} show the measured values.  Errors in the last column do
not include errors on $N_\gamma ^{sim}$. The reference value of
quantum efficiency given by the manufacturer is
$\epsilon_q^{DUV}(150nm) = 1.07$ and $\epsilon_q^{DUV}(175nm) = 1.22$.
Our data show good agreement with the values quoted by the
manufacturer \cite{advphot}.

\begin{table}[h!]
\centering
\begin{tabular}{|c|c|c|c|c|c|c|}
\hline
  $p_{Kr}$  & $T_{Kr}$  & Shaping & Gain & $Q_s$  &  $N_\gamma ^{sim}$ & $\frac{Q_S}{e\cdot N_\gamma^{sim}\cdot G} $\\
(atm) &(${}^\circ$K) & ($\mu$s) && (fC)&&\\
\hline
\hline
  0.629 & 285.0 & 10  & $26.1\pm 1.8$ &  $15.6\pm 0.6$ & 3844 &  $0.97\pm 0.08$ \\
  0.600 & 284.4 & 10  & $28.1\pm 2.1$ &  $15.8\pm 0.6$ & 3613 & $ 0.97\pm 0.08$  \\
  0.573 & 284.0 & 10  & $25.1\pm 1.6$ &  $15.5\pm 0.6$ & 3402 & $ 1.13\pm 0.08$  \\
  0.551 & 283.9 & 10  & $25.3\pm 1.6$ &  $14.3\pm 0.5$ & 3246 & $ 1.08\pm 0.08$  \\
  0.514 & 283.9 & 10  & $27.3\pm 1.9$ &  $13.8\pm 0.5$ & 2960 & $ 1.06\pm 0.08$  \\
\hline
Kr & DUV-enh & & & & & $1.04\pm 0.08 $ \\
\hline
\end{tabular}
\vspace{0.5cm}
\caption{\label{tab:krxe}The gain-independent signal normalized to the
expected number of DUV photons in krypton (See text).}
\end{table} 

\begin{table}[h!]

\centering
\begin{tabular}{|c|c|c|c|c|c|c|}
\hline
  $p_{Xe}$  & $T_{Xe}$  & Shaping & Gain & $Q_s$  &  $N_\gamma ^{sim}$ & $\frac{Q_S}{e\cdot N_\gamma^{sim}\cdot G} $\\
(atm) &(${}^\circ$K) & ($\mu$s) && (fC)&&\\
\hline
\hline
 0.430 & 286.4 & 10  & $26.1\pm 1.8$ &  $20.6\pm 0.8$ & 3786 & $ 1.29\pm 0.10$ \\
 0.407 & 285.3 & 10  & $25.6\pm 1.7$ &  $19.1\pm 0.7$ & 3551 & $ 1.31\pm 0.10 $\\
 0.379 & 284.9 & 10  & $25.6\pm 1.7$ &  $18.0\pm 0.7$ & 3256 &  $1.35 \pm 0.10$\\
 0.372 & 286.0 & 10  & $24.6\pm 1.6$ &  $16.0\pm 0.6$ & 3163 &  $1.28\pm 0.10$ \\
\hline
Xe &DUV-enh & & & & & $1.3\pm 0.1$ \\
\hline
\end{tabular}
\vspace{0.5cm}
\caption{\label{tab:xe}The gain-independent signal normalized to the
expected number of DUV photons in xenon (See text).}
\end{table}

\begin{table}[h!]
\centering
\begin{tabular}{|c|c|c|c|c|c|c|}
\hline
 $p_{Ar}$  & $T_{Ar}$  & Shaping & Gain & $Q_s$  &  $N_\gamma ^{sim}$ & $\frac{Q_S}{e\cdot N_\gamma^{sim}\cdot G} $\\
(atm) &(${}^\circ$K) & ($\mu$s) && (fC)&&\\
\hline
  0.835 & 278.7 & 5  & $87 \pm 6$ &  $32.7\pm 1.2$ & 3337 & $ 0.70\pm 0.05$ \\
  0.808 & 278.4 & 5  & $96\pm 8$ & $32.9\pm 1.2$  & 3181 & $ 0.67\pm 0.06$ \\
  0.783 & 278.4 & 5  & $63\pm 5$ & $18.5\pm 0.7$ & 3037 & $ 0.60\pm 0.05$ \\
  0.757 & 278.4 & 5  & $52\pm 4$ & $14.5\pm 0.5$  & 2895 & $ 0.60\pm 0.05$\\ 
  0.727 & 278.2 & 5  & $38\pm 3$ & $9.5\pm 0.4$ & 2738 & $ 0.56\pm 0.04 $\\
  0.699 & 278.2 & 5  & $39\pm 3$ & $8.5\pm 0.3$  & 2598 & $ 0.53\pm 0.04$\\
\hline 
\hline
  0.830 & 282.5 & 10  & $69\pm 5$ & $ 25.6\pm 1.0$ & 3238 & $ 0.72\pm 0.05$ \\
  0.777 & 281.5 & 10  & $67\pm 5$ &  $24.5\pm 0.9$ & 2956 &  $0.77\pm 0.06$ \\
  0.750 & 281.4 & 10  & $57\pm 4$ & $ 17.7\pm 0.7$ & 2815 & $ 0.69\pm 0.05$ \\
  0.719 & 281.3 & 10  & $57\pm 4$ &  $16.7\pm 0.5$ & 2659 & $ 0.69\pm 0.05$ \\
  0.678 & 281.2 & 10  & $53\pm 4$ & $ 15.5\pm 0.7$ & 2461 & $ 0.74\pm 0.06$ \\
\hline
Ar &DUV-enh & & & & & $0.72\pm 0.06$ \\
\hline
\end{tabular}
\vspace{0.5cm}
\caption{\label{tab:ar}The gain-independent signal normalized to the
expected number of DUV photons in argon (See text). Measurements
performed using DUV-enhanced LAAPD. The last line gives the average
value of the 10 $\mu$s measurements. }
\end{table} 

Following the same procedure, measurements were made in argon. The
measured values are listed in Table~\ref{tab:ar}. 

\begin{table}[h!]

\centering
\begin{tabular}{|c|c|c|c|c|c|c|}
\hline
 $p_{Ar}$  & $T_{Ar}$  & Shaping & Gain & $Q_s$  &  $N_\gamma ^{sim}$ & $\frac{Q_S}{e\cdot N_\gamma^{sim}\cdot G} $\\
(atm) &(${}^\circ$K) & ($\mu$s) && (fC)&&\\
\hline
 
 0.835 & 275.9 & 1  & 79.9 & 10.2  & 3388 &  0.23 \\
 0.820 & 276.7 & 1  & 77.6 & 9.39  & 3281 &  0.23 \\
 0.806 & 276.7 & 1  & 80.0 & 9.98  & 3197 &  0.24 \\
\hline
Ar& Red/IR-enh  &  &  &  &  & 0.233\\
\hline
0.862 & 278.7 & 1  & 90.4 & 7.2  & 3505 &  0.14 \\
0.847 & 279.2 & 1  & 159.9 & 13.6  & 3400 &  0.16 \\
0.830 & 279.1 & 1  & 168.5 & 14.5  & 3303 &  0.16 \\
\hline
Ar & Red/IR+Filter&&&&& 0.153\\
\hline
\end{tabular}
\vspace{0.5cm}
\caption{\label{tab:arir}Measurements performed in argon with the red/IR-enhanced
LAAPD. The lower data is taken with a UV absorbing foil in front (See text).}
\end{table} 

\section{Signal Length and Shaping Time}

\begin{figure} 
\includegraphics[width=\textwidth]{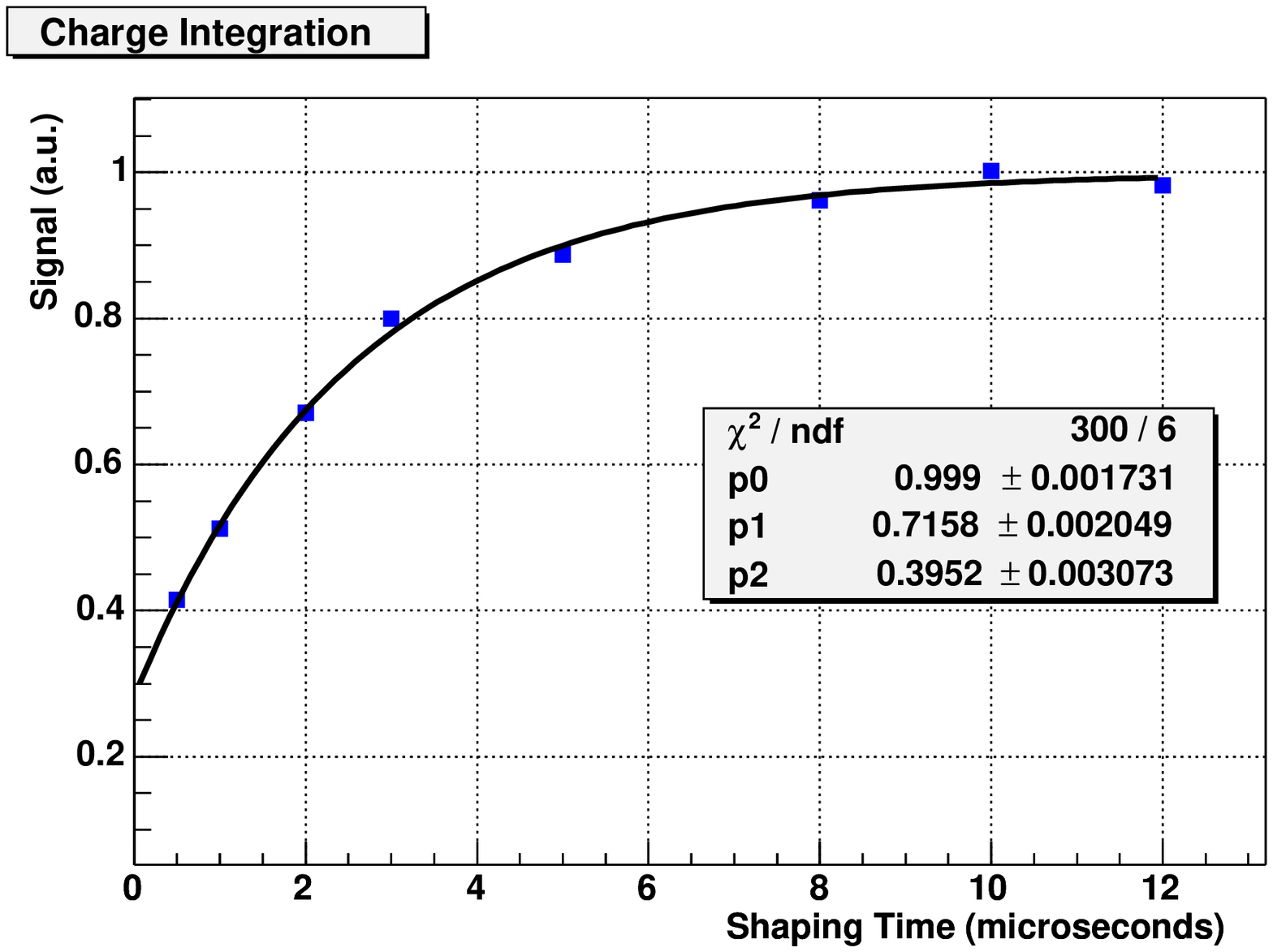} 
\caption{\label{sha} Long shaping times are necessary to integrate full
charge. Plot shows relative signal size for different shaping
times. Data taken in argon at 0.8 atm and 280 degrees K. Plotted
errors are solely statistical.}
\end{figure}

The mechanism of DUV light emission in low pressure noble gas occurs
on a time scale of several microseconds, see~\cite{moerman}.
Therefore, not all charge is integrated when running measurements at
shaping times that optimize energy resolution. Series of data (not
listed in Table \ref{tab:ar}) were taken at different shaping times. The
results are shown in Figure \ref{sha}, where the signal height
is plotted in arbitrary units as a function of the shaping time
used. The data was fitted to a function
\begin{equation}
\label{eq:c3}
S(\tau)=p_0 - p_1\cdot \exp(-p_2\cdot \tau)
\end{equation}
where $\tau$ is the shaping time in microseconds. The fitted decay
frequency of $p_2=0.395 \mu s^{-1}$ is comparable to the value
$p_2=0.2 + 0.12 \cdot p = 0.297\mu s^{-1}$ obtained by
\cite{moerman}. We conclude for argon pressures above .678
atmospheres, the use of 10 $\mu$s shaping time guarantees the
integration of at least 94$\%$ of the full signal.

\begin{figure} 
\includegraphics[width=\textwidth]{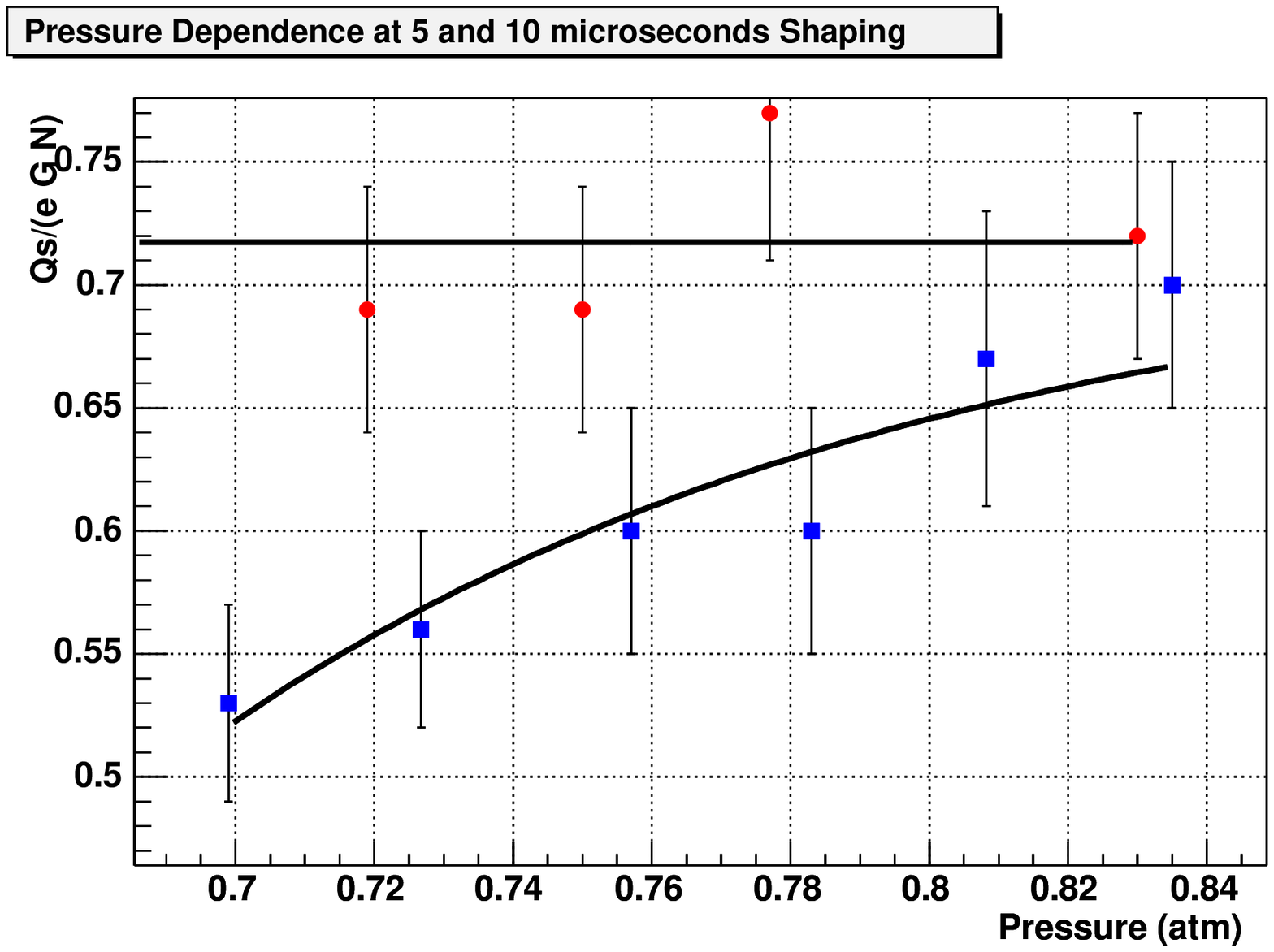} 
\caption{\label{arp} The plot shows a systematic pressure dependence
of data taken at short shaping times, 5$\mu$s in the case of the data
plotted by the square markers. The data taken at 10$\mu$s shaping
time, plotted as circles, shows no systematic pressure dependence. The
error bars do not include errors on $ N_\gamma $.}
\end{figure}

This fact is confirmed by the data taken. The series taken at 5 $\mu$s
shaping time shows a clear pressure dependence, see Table \ref{tab:ar}
and Figure \ref{arp}. The series taken with 10 $\mu$s shaping time
shows no systematic pressure dependence.

The energy resolution is optimized at a finite value of shaping
time. For measurements performed at an APD temperature of
$~10^\circ$C, the best energy resolution of $\sigma / \bar{x} = 0.085$
is achieved with 2 $\mu$s shaping.  As the APD temperature is reduced,
dark current and consequently parallel noise decreases, reducing the loss
of resolution at long shaping times.

\section{Non-UV Contribution}
\label{IRC}

Non-UV emission is assumed to be emitted from atomic transitions
before dimer formation. 

For argon, measurements were performed to allow the subtraction of the
possible non-UV contribution. This was done by replacing the
DUV-enhanced LAAPD by a red/IR-enhanced LAAPD from the same
manufacturer.  The measurements are shown in Table~\ref{tab:arir}.
Note that $N_\gamma ^{sim}$ in the table represents the number of
expected DUV photons. The device, however, is not primarily sensitive
to this wavelength. Therefore, the signal normalized to the expected
number of DUV photons should not be interpreted directly as a quantum
efficiency. Rather, this result is to be interpreted as a significant
contribution of non-UV photons to argon scintillation light.

Because of its direct production mechanism, non-UV emission should occur on a
faster time scale than DUV emission. Our non-UV measurements were
performed at 1 $\mu$s shaping time. Note from Table \ref{tab:arir}
that no systematic pressure dependence of the non-UV signal seems to
be present. This observation is consistent with the attribution of
non-UV emission to transitions between excited atomic states which are
precursors of dimer (DUV) emission.

The red/IR enhanced LAAPD has an $SiO_2$ anti-reflective coating. In
 general, $SiO_2$ is opaque to UV light. It is not clear to what
 extent such an effect attenuates 128nm light passing through this 150
 nm thin coating. We do not a priori exclude a residual sensitivity of
 the red/IR enhanced device to DUV light. 

To reduce this ambiguity, a plastic foil of 0.1 mm thickness was
employed, mounted in front of the red/IR-enhanced device. It can be
assumed that this foil has no transmittance in the DUV range, and a
finite transmittance in the Red/IR range. The foil caused a signal
reduction of 34$\%$, see Table \ref{tab:arir} . This leads us to
conclude that at least 65$\%$ of the unfiltered signal detected with
the red/IR-enhanced LAAPD can be attributed to non-UV photons.

From now on, we use the super-index ${IR}$ when referring to the
red/IR-enhanced APD.  Likewise, the super-index ${DUV}$ denotes the
DUV-enhanced device. The number of non-UV photons is written as
$N^{IR}$. Values of quantum efficiency always are a function of
the wavelength given in parenthesis.

Using the data listed in Table \ref{tab:arir}, we can give a strict
lower limit for the branching ratio $N^{IR}/N^{sim}_\gamma $ of the
emission of non-UV photons in argon. The number of non-UV photons impinging on the Red/IR-enhanced LAAPD relates to the detected charge signal in linear dependence of gain and quantum efficiency:
\begin{equation}
N^{IR}=\frac{Q_S}{e\cdot \epsilon_q^{IR}(\lambda) \cdot G}
\end{equation}
The expression is minimized by the maximum quantum efficiency of the Red/IR-enhanced APD, giving the expression
\begin{equation}
\label{eq:ir}
\left(\frac{Q_S}{e\cdot N_\gamma^{sim} \cdot
G}\right)_{Red/IR+Filter}\cdot \frac{1}{\max_{270\le\lambda\le1050}
\epsilon_q^{IR}(\lambda)} = 0.18 \le \frac{N^{IR}}{N_\gamma^{sim}} ,
\end{equation}
where $ \max_{\lambda}( \epsilon_q^{IR})=0.85 $ was used. By
comparison with $W_\gamma^{DUV}$ (See Section~\ref{sec:nobgas}), the
obtained value can be translated into
\begin{equation}
 W_\gamma^{IR} \le 378 \, \rm  eV,
\end{equation}
 a strict upper limit for the average amount of energy needed to
produce a non-UV photon in argon gas around this pressure.

With the data in Table \ref{tab:arir}, the quantum efficiency of the
DUV-enhanced LAAPD for radiation at 128 nm can be calculated more
precisely 
\begin{equation}
\epsilon_q^{DUV}(128 nm)= \epsilon_{q0}^{DUV}(128 nm)-  N^{IR}(\lambda)\cdot\epsilon_q^{DUV}(\lambda) 
\end{equation}
where $\epsilon_{q0}^{DUV}(128 nm)$ refers to the left side of Equation \ref{eq:eqcal} and the last term is the correction for the sensitivity of the DUV-enhanced device to any non-UV photons of wavelength $\lambda$ emitted.

As a strict lower limit, 
\begin{equation}
 \left(\frac{Q_S }{e\cdot N_\gamma^{sim}\cdot G}\right)_{DUV}-
\left(\frac{Q_S }{e\cdot N_\gamma^{sim}\cdot G}\right)_{IR}\cdot \max_{270\le \lambda \le 1050} 
\frac{\epsilon_q^{DUV}(\lambda)}{\epsilon_q^{IR}(\lambda)}
\le \epsilon_q^{DUV}(128nm),
\end{equation}
and, as an upper limit
\begin{equation}
 \left(\frac{Q_S }{e\cdot N_\gamma^{sim}\cdot G}\right)_{DUV}-
\left(\frac{Q_S }{e\cdot N_\gamma^{sim}\cdot G}\right)_{IR+Filter}\cdot \min_{270\le \lambda \le 1050} 
\frac{\epsilon_q^{DUV}(\lambda)}{\epsilon_q^{IR}(\lambda)}
\ge \epsilon_q^{DUV}(128nm).
\end{equation} 
can be given. Thus, 
\begin{equation}
0.42 \le \epsilon_q^{DUV}(128nm) \le 0.73
\end{equation}
where the upper limit is not rigorous as quenching effects due to gas
impurities have not been accounted for.

If argon non-UV emission is centered around 940 nm as measured by
\cite{bressi}, this would result in an in-between value of
$\epsilon_q^{DUV}(128nm)\approx 0.58 $. Note that over a large region
of the IR spectrum, the ratio of $\epsilon_q^{DUV}/\epsilon_q^{IR}$ is
relatively constant, making $\epsilon_q^{DUV}(128nm)$ relatively
insensitive to the exact wavelength of peak IR emission. 

In xenon and krypton, the non-UV contribution was not quantitatively
 measured.

\begin{figure}[tb] 
\includegraphics[width=\textwidth]{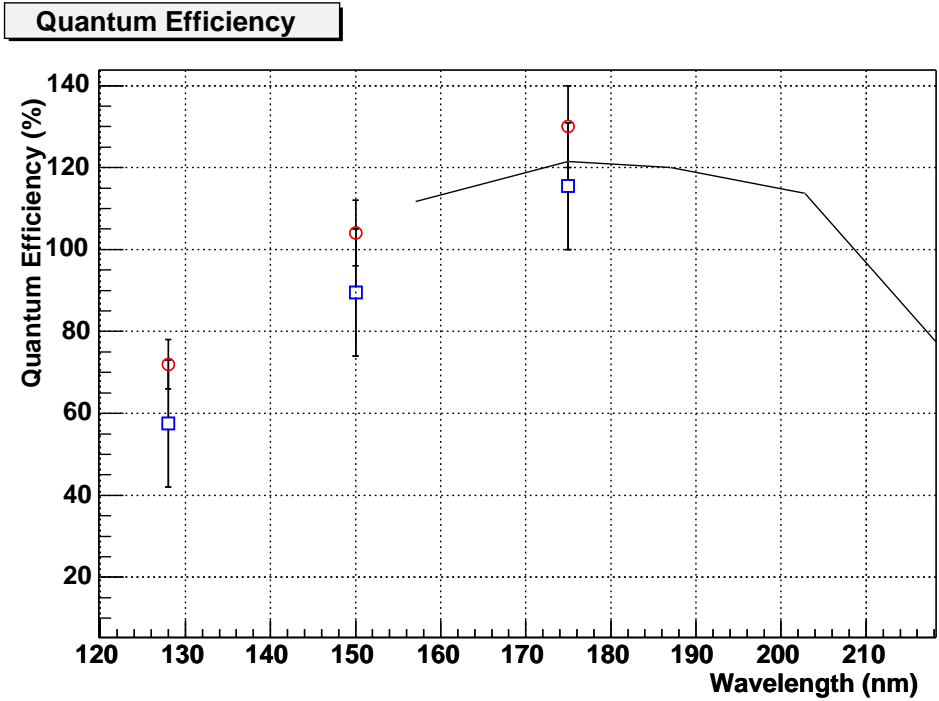} 
\caption{\label{qeffdec} The quantum efficiency of the DUV-enhanced
LAAPD: The continuous curve depicts the values given by the
manufacturer, see Figure \ref{APIqe}. The circles represent the values
from Tables \ref{tab:ar}, \ref{tab:krxe}, and \ref{tab:xe} at argon,
krypton, and xenon emission wavelengths. The square markers include the
correction for non-UV components (see Section~\ref{IRC}).}
\end{figure}

\section{Conclusion}

Deep ultraviolet light can be detected by LAAPDs with significantly
higher quantum efficiency than by conventional means such as
photomultipliers.

Our data is consistent with the fact that argon gas at low pressure
emits a significant amount of non-UV light. A strict upper limit for
the average energy necessary to produce a non-UV photon in argon gas
at a pressure below 1 atm is given. A lower limit cannot be given
since our device is not sensitive throughout the IR full spectrum.

The given limits for the $\epsilon_q^{DUV}(128nm)$  apply for argon
non-UV emission down to 270 nm. Since impurity quenching was not
considered, the lower limit of $42\%$ is strict while the upper limit
is not. For non-UV emission centered around 940 nm
$\epsilon_q^{DUV}(128nm)\approx 58\% $ is obtained.

Figure~\ref{qeffdec} summarizes the obtained results for argon,
krypton, and xenon. Our measurements are consistent with the manufacturer's data 
where it is available. The non-UV correction was
only measured for argon.  For xenon and krypton this correction is of illustrative
nature only. Error bars do not include inaccuracies of
$N^{sim}_\gamma$, nor do they include quenching effects which lead to
efficiency under-estimation. Our results tend to underestimate the
quantum efficiency. This may be attributed to the presence of UV
attenuating impurities, whose presence are not considered in the
calculation. If this is the case, the effect can be expected to be
more severe at shorter wavelengths. In this sense, our measurement of
the quantum efficiency at 128 nm is to be seen as a lower limit.

\section*{Acknowledgments}
We are indebted to the INFN Padova group who has cordially lent us the
readout electronics necessary for the measurements. In particular, we
thank Sandro Centro (INFN Padova) for his support. We acknowledge
Francesco Pietropaolo and Pio Picchi who indirectly
contributed to this study through useful discussions.

This work was supported by ETH/Z\"urich and Swiss National Science Foundation.



\end{document}